\documentclass[aps,prl,amsmath,amssymb,twocolumn,10pt,]{revtex4-1}

\usepackage{amssymb,amsfonts,amsmath}

\usepackage{graphicx}
\usepackage{epstopdf}
\usepackage[extension=xxx]{hyperref}
\newcommand{\unit}[1]{\ensuremath{\, \mathrm{#1}}}

\usepackage{color}

\def\be{\begin{equation}}   \def\ee{\end{equation}}
\def\eq#1{{Eq~(\ref{#1})}}    \def\fig#1{{Fig.\ref{#1}}}

\def\ff{{f_{\rm f}}}
\def\ron{{r_{\rm on}}}  \def\koff{{k_{\rm off}}}   \def\kon{{k_{\rm on}}}   \def\koffo{{k_{\rm off}^0}}

\begin{document}

\title{Stick-Slip model for actin-driven cell protrusions, cell polarisation and crawling.}
\author{Pierre Sens}
\affiliation{Institut Curie, PSL, 26 Rue d'Ulm, 75005, Paris}
\affiliation{Sorbonne Universit\'es, UPMC Univ Paris 06, 75005, Paris, France}


\begin{abstract}
Cell crawling requires the generation of intracellular forces by the cytoskeleton and their transmission to an extracellular substrate through specific adhesion molecules. Crawling cells show many features of excitable systems, such as spontaneous symmetry breaking and crawling in the absence of external cues, and periodic and propagating waves of activity. Mechanical instabilities in the active cytoskeleton network and feedback loops in the biochemical network of activators and repressors of cytoskeleton dynamics have been invoked to explain these dynamical features.
Here, we show that the interplay between the dynamics of cell-substrate adhesion and linear cellular mechanics is sufficient to reproduce many non-linear dynamical patterns observed in spreading and crawling cells. 
Using an analytical formalism of the molecular clutch model of cell adhesion, regulated by local mechanical forces, we show that cellular traction forces exhibit a stick-slip dynamics resulting in periodic waves of protrusion/retraction and propagating  waves along the cell edge. 
This can explain spontaneous symmetry breaking and polarisation of spreading cells, leading to steady crawling or bipedal motion, and bistability, where persistent cell motion requires a sufficiently strong transient external stimulus. 
The model also highlight the role of membrane tension in providing the long-range mechanical communication across the cell required for symmetry breaking.
\end{abstract}

\keywords{cell motility | symmetry breaking | stick-slip | Membrane tension |}

\date{\today}
 \maketitle



Cell crawling is ubiquitous in many  biological processes from development to cancer. It is inherently a problem of mechanics, in which forces generated by the cytoskeleton are transmitted to the environment through transient adhesion to allow for cell translocation \cite{lauffenburger:1996}. 
The cytoskeleton is a highly dynamical active gel able to exert pushing forces through the polymerisation of actin filaments and contractile forces through the interaction between actin and myosin motors. 
In a schematic description of cell crawling, the protrusion of the cell front is driven by actin polymerisation while acto-myosin contraction retracts the rear of the cell \cite{ananthakrishnan:2007}. 
In their physiological context, cells often polarise and crawl in response to external cues, such as gradient of chemotractants or of mechanical properties of their environment \cite{graziano:2014,charras:2014}. However, many cells also crawl as a result of spontaneous symmetry breaking, and exhibit periodic and/or propagating waves of activity \cite{allard:2012}. Even cell fragments devoid of nucleus show spontaneous symmetry breaking and bistability, and can be driven into a persistent motile state by transient mechanical stimulii \cite{Verkhovsky:1999}.

These non-linear features call for a description of motile cells as self-organised systems in which feedback loops lead to dynamical phase transitions \cite{mogilner:2009b,holmes:2012,danuser:2013}. Many such descriptions have been proposed, most of which focus on the behaviour of the cytoskeleton itself. 
One class of models, which include bistability, polarisation and wave propagation, is based on the existence of feedback loops within the biochemical network of proteins regulating cytoskeletal activity, such as Rac GTPases which activates actin polymerization and protrusion, or Rho GTPases which activates actomyosin contractility  \cite{sohrmann:2003,weiner:2007,machacek:2009}. This includes possible mechanical feedback, for instance through modulations of the cell membrane tension \cite{houk:2012,dizmunoz:2013}. 
Another class of models focuses on the  mechanics of the cytoskeleton, an active viscoelastic gel made of polar filaments which can spontaneously form asters and vortices \cite{kruse:2004}. Symmetry breaking \cite{callan:2008} and spontaneous motility \cite{blanch:2013} can be obtained by coupling filament orientation to the cell boundary, and waves can arise from the reaction-diffusion dynamics of actin nucleators and inhibitors \cite{doubrovinski:2011}. Modulations of the myosin distribution by the actin flow has also extensively been studied, and can lead to instabilities \cite{bois:2011} and spontaneous polarisation and motion \cite{hawkins:2011,ziebert:2012,tjhung:2012,ruprecht:2015,Maiuri:2015}. 
In fast moving, crescent-shaped cells such as keratocytes and cell fragments, the actin cytoskeleton often forms a branched network at the cell front and contractile bundles enriched in myosin at the back. A switch-like transition between these two structures has been described using phenomenological models \cite{kozlov:2007,lomakin:2015}. Finally, many models have also addressed the shape, dynamics and speed of motile cells through feedback between shape and the rate of actin polymerisation and depolymerisation \cite{lee:1993,keren:2008,ofer:2011,barnhart:2011,barnhart:2017}.

The models above generally treat force transmission with the substrate as a simple linear friction. The present work focusses on the non-linear dynamics of cell-substrate adhesion. More specifically, we concentrate on so-called mesenchymal cell motility on flat substrate, where a thin protrusion called the lamellipodium forms the leading edge of spreading and crawling cells, powered by actin polymerisation \cite{ridley:2011}. Polymerisation is often offset by a retrograde flow of actin away from the cell edge, driven by acto-myosin contraction and the cell membrane tension.  According to the ''molecular clutch'' model \cite{mitchison:1988,giannone:2009}, these retrograde forces are balanced by frictional traction forces resulting from the transient linkage between actin filaments and the substrate through the binding and unbinding of adhesion molecules such as integrins. 
This linkage involves a myriad of regulatory proteins \cite{zaidelbar:2003}, many of which are mechanosensitive \cite{bershadsky:2006,evans:2007,giannone:2009,ladoux:2012}. The lifetime of individual bonds can decrease (slip-bonds \cite{merkel:1999,jiang:2003}) or increase (catch-bonds \cite{marshall:2003,kong:2009,thomas:2008,delrio:2009}) under force, which can lead to a stick-slip dynamics \cite{filipov:2004,chan:2008}, as detailed below. This provides a natural explanation for the existence of two (slipping and gripping) states of actin flow dynamics \cite{jurado:2005,hu:2007}), for the biphasic relationship between actin flow and traction force \cite{gardel:2008,aratyn:2010,li:2010,craig:2015}, and for the traction force dependence on substrate stiffness \cite{pio:sliding,bangasser:2013} 

A stick-slip transition has been extensively discussed in the context of cell protrusion and motility, either using a discontinuous version of the stick-slip transition where adhesion sites break beyond a critical force \cite{wolgemuth:2005,barnhart:2010,shemesh:2012,loosley:2012,ziebert:2013,barnhart:2015} or quantitatively accounting for the dynamics of the transition in a population of stochastic bonds \cite{chan:2008,shao:2012,prahl:2020}. However, we are still in need of a simple analytical understanding of the interplay between the stick-slip dynamics and other dynamical processes within the cell, which has thus far mostly been studied through computer simulations. The present work offers such a description within a linear cell mechanics framework based on simple elastic or visco-elastic constitutive relationships. It mostly focusses on one-dimensional (1D) cells, although lateral propagating waves are also discussed. Even this simplified model displays a rich dynamical behaviour, including protrusion/retraction waves, spontaneous polarisation or bistability, and unsteady motion. This can be used has a firm basis to understand more complicated systems with multiple interacting feedback loops.

\section{Model and Results.}
\begin{figure}[t]
\centerline{\includegraphics
[width=1\linewidth]{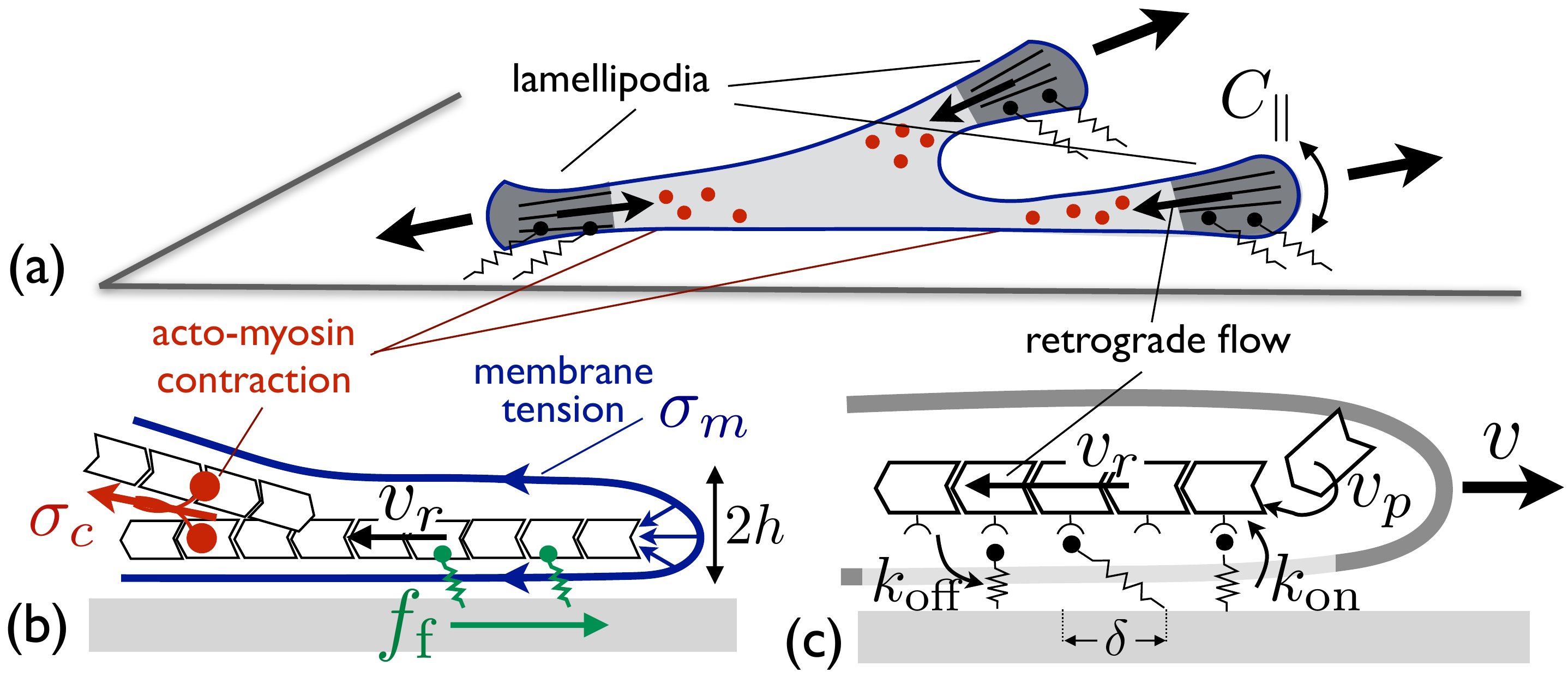}}
\caption{{\bf (a)} Sketch of a 2D cell with three protrusions, showing the local cell edge curvature $C_\|$. {\bf (b)}  Force balance at the cell edge between the friction force $\ff$ and the ``retrograde force'' from membrane tension $\sigma_m$ (blue arrows at the tip show the local Laplace pressure) and acto-myosin contraction $\sigma_c$.
{\bf (c)} Binding/unbinding cycle of adhesive linkers on a moving actin filament polymerizing near the cell edge. 
\label{fig1_sketch}}
\end{figure}

Actin filaments polymerising against the cell membrane and acto-myosin contraction create an actin retrograde flow away from the cell edge. This flow correlates with high substrate traction stress and is often concentrated in the lamellipodium, near the cell edges  \cite{ponti:2004,gardel:2008} (\fig{fig1_sketch}a). 
According to  the molecular  clutch model, traction stress is akin to a friction force exerted by adhesion molecules  transiently bound to moving filaments  (\fig{fig1_sketch}c).
The retrograde velocity is fixed by the force balance (sketched in \fig{fig1_sketch}b) between this friction force and the ``retrograde force'' from the cytoskeleton tension $\sigma_c$ - mostly due to the acto-myosin contraction - and the membrane tension $\sigma_m$ \cite{pio:motil_tension}. The latter force is the membrane Laplace pressure integrated over the lamellipodium thickness $2h$: $2h\sigma_m C$, where the total curvature $C=1/h+C_\|$ is the sum of the curvature along the lamellipodium height $1/h$ and the curvature along the cell edge $C_\|$ (see \fig{fig1_sketch}a). Assuming a uniform retrograde flow velocity $v_r$ over the lamellipodium for simplicity and calling  $\ff(v_r)$ the friction force per unit  length along the cell edge, integrated over the lamellipodium depth, the local force balance reads:
\be
\ff(v_r)=\sigma_c+2\sigma_m(1+hC_\|).
\label{forcebal}
\ee

\subsection{Mechanosensitive adhesion and stick-slip dynamics}
The traction force is mediated by protein linkers binding to and unbinding from actin filaments with rates $\kon$ and $\koff$ (\fig{fig1_sketch}c). Both rates might depend on the force $f_b$ felt by a given linker. In the following, we concentrate on mechano-sensitive off-rate and define the dimensionless, force-dependent off-rate  $r(f_b)=\koff/\koffo$, where $\koffo$ is the off-rate under zero force. Similarly, the dimensionless on-rate is $\ron=\kon/\koffo$ and the dimensionless time is $\bar t=\koffo t$.
The stochastic friction force is proportional to the fraction of available linkers attached to actin filaments at a given time, called $n$, multiplied by their average extension  $\delta\sim v_r/r$  (\fig{fig1_sketch}c and Supplementary Section S2).
The kinetic equation for the fraction of bound linkers and the total substrate friction force read \cite{schallamach:1963,tawada:1991,chan:2008,sabass:2010}
\be
\frac{\partial n}{\partial \bar t}=\ron -(\ron+r)n\quad {\rm and}\quad  \ff(v_r)=\left(\alpha_0+\alpha_1 \frac{n}{r}\right)v_r,
\label{ff1}
\ee
where $\alpha_1$ is the coefficient of stochastic friction and  $\alpha_0$ is a bare friction coefficient characterising other (linear) viscous dissipation  between the actin flow and the substrate. The former can be expressed in terms of the linkers density $\rho$ and stiffness $k_b$ as $\alpha_1=\rho l_{\rm l} k_b/\koffo$, where $l_{\rm l}$ is the lamellipodium width.

\begin{figure}[b]
\centerline{\includegraphics
[width=1\linewidth]{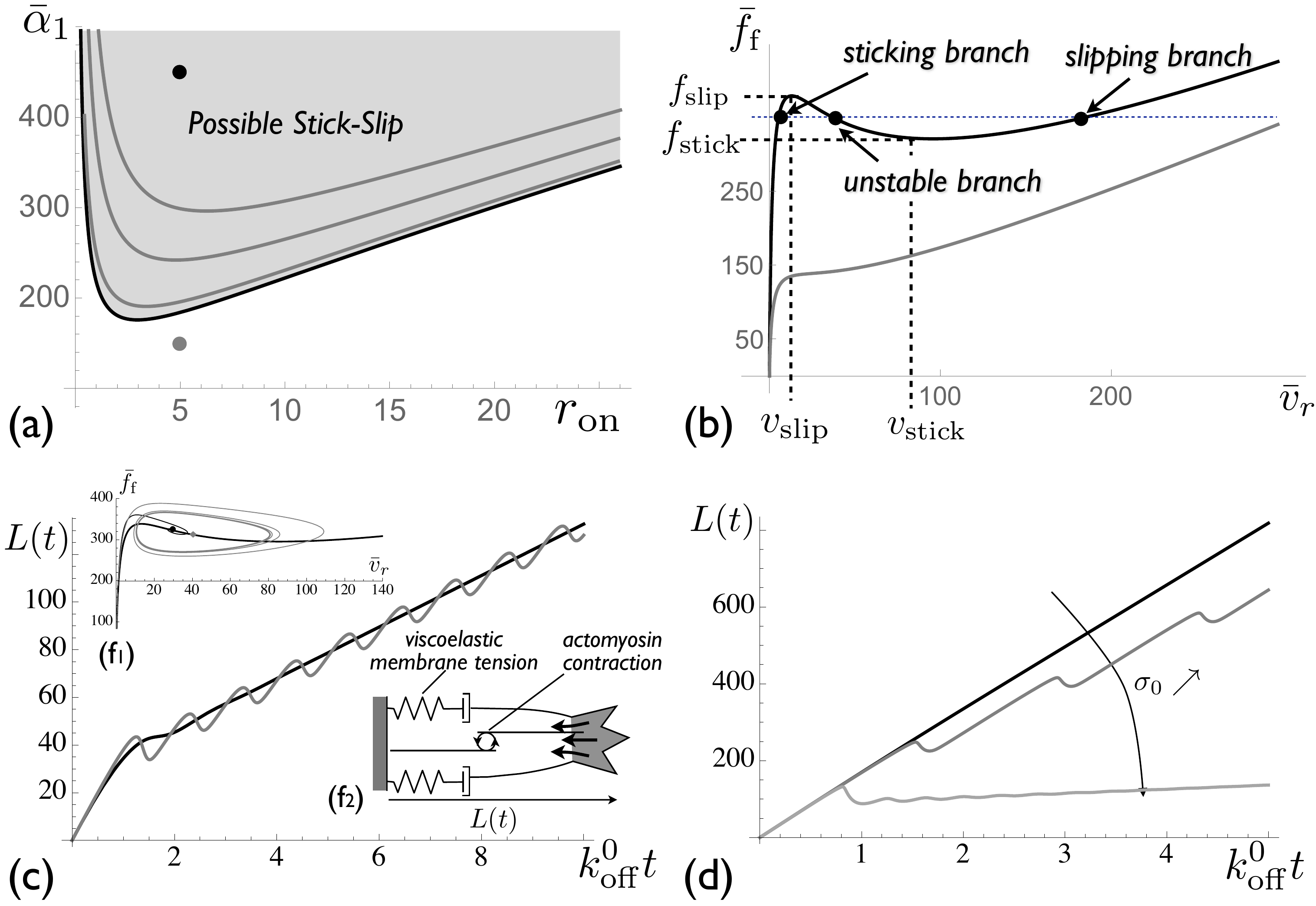}}
\caption{\small {\bf (a)} Possible stick-slip region in the parameter space $\{\ron,\bar\alpha_1=\alpha_1/\alpha_0\}$ for an elastic protrusion of dimensionless stiffness $\bar k_\sigma=k_\sigma/(\alpha_0 \koffo)$. The shaded region corresponds to a non-monotonous friction force. Black line: stick-slip boundary for infinitely fast linker kinetics ($\bar k_\sigma\rightarrow0$), gray lines: boundaries for $\bar k_\sigma=1,\ 5$ and $10$ (\eq{stab_elast}). 
{\bf (b)} Example of force/velocity relationship in and out of the stick-slip region (for $\ron=5$ and $\bar\alpha_1=150$ (gray) or $450$ (black)). 
Characteristic values of the retrograde velocity ($\bar v_{\rm slip}$ and $\bar v_{\rm stick}$) and force ($f_{\rm slip}$ and $f_{\rm stick}$) are defined.
{\bf (c,d)} Growth of a 1D viscoelastic protrusion (sketched) for $\bar\alpha_1=400$ and $\ron=5$.
{\bf (c)} Protrusion length with time for two viscoelastic relaxation times and the same average speed, black: $\bar\tau_\sigma=0.2$ and $\bar\sigma_0=300$, gray:  
$\bar\tau_\sigma=1$ and $\bar\sigma_0=214$ (with $r_p=15$). 
{\bf (d)} Protrusion length with time for different homeostatic tension (with $\bar\tau_\sigma=0.5$ and $r_p=45$).
\label{fig2_protrusion}}
\end{figure}

For constant  binding and unbinding rates, the stochastic friction force is linear with the retrograde velocity~\cite{tawada:1991}. 
For slip-bonds, generic thermodynamic arguments suggest that the off-rate increases exponentially with the force per bond  $f_b$: $r=e^{|f_b|/f_b^*}$,  where   $f_b^*$ is a characteristic molecular force scale \cite{bell:1978,evans:2007}. The retrograde velocity can thus be directly related to the off-rate:  $v_r=v_\beta r\log r$, where  $v_\beta= \koffo f_b^*/k_b$ is a velocity scale  characterising the mechano-sensitivity of unbinding. The steady-state friction force is non-linear with the retrograde velocity, and reads:
\be
\ff^*=\alpha_0 v_\beta \left(r+\bar\alpha_1\frac{\ron }{\ron+r}\right)\log{r}\ \ ;\ \ v_r=v_\beta r\log{r}
\label{ss}
\ee
where $\bar\alpha_1 \equiv \alpha_1/\alpha_0$. This defines a regime of high friction for small retrograde velocity when most linkers are bound, and low friction dominated by the viscous drag $\alpha_0$  for large  velocity when most linkers are unbound. Remarkably, the force-velocity relationship \eq{ss} is non-monotonous for a broad range of parameters, with an abrupt transition between the two regimes (\fig{fig2_protrusion}a,b), equivalent to a stick-slip transition \cite{filipov:2004}. This occurs when the ratio of stochastic to viscous friction $\bar\alpha_1$ is large, which is expected for crawling cells.  
In this  regime, we may define characteristic values of the force and retrograde velocity (\fig{fig2_protrusion}b). The high friction regime exists for small forces $\ff<f_{\rm slip}$ and small retrograde velocity $v_r<v_{\rm slip}$, and the low friction regime  exists for large force $\ff>f_{\rm stick}$ and large velocity $v_r>v_{\rm stick}$.

Catch-bond effects lead to a biphasic dependence of the unbinding rate, decreasing under small force and increasing under larger force \cite{
novikova:2013,bangasser:2013}. This strongly amplifies the stick-slip behaviour (Supplementary Section S.2 and Fig.S1).
This could be of strong physiological relevance, but does not qualitatively change the results described below and the following analysis concentrates on slip-bonds.

\subsection{Dynamics of protrusion powered by actin polymerisation} 
We first focus on the dynamics of a unidimensional cellular protrusion, or a cell front moving at uniform velocity ($C_\|=0$ in \eq{forcebal}). The protrusion length $L(t)$ grows through a balance between actin polymerisation at a velocity $v_p$ at the protrusion tip and actin retrograde flow: $\partial_t L=v_p-v_r$ (\fig{fig1_sketch}c). The retrograde force from membrane tension and acto-myosin contraction may evolve in time. The cell membrane tension has a visco-elastic behaviour. It increases (up to four-fold) following the formation of cell protrusions or cell spreading and relaxes at longer times \cite{keren:2011b,gauthier:2011,houk:2012}, through the flattening of membrane invaginations such as caveolae \cite{pio_regul,sinha:2011} or via endocytosis and exocytosis \cite{bretscher:1998,gauthier:2012}. Disregarding spatio-temporal modulation of acto-myosin activity,  the total tension $\sigma=\sigma_c+2\sigma_m$ is described as a generic Maxwell fluid with an  short-time elastic behaviour characterised by an effective stiffness $k_\sigma$ and a  long-time viscous relaxation toward an homeostatic tension $\sigma_0$ with a relaxation time $\tau_\sigma$:  
\be
\partial_t \sigma+\frac{\sigma-\sigma_0}{\tau_\sigma}=k_\sigma\partial_t L=k_\sigma(v_p-v_\beta r\log{r}),
\label{maxwell}
\ee

\underline{\em Elastic protrusions.}  In the elastic regime ($\tau_\sigma\rightarrow\infty$), the retrograde velocity balances the polymerisation velocity for a stationary protrusion length $L^*$ satisfying $\sigma(L^*)=\ff^*(r_p)$, where $r_p$ is the stationary off-rate: $v_\beta r_p\log{r_p}=v_p$. Linear stability analysis of a perturbation around the steady-state: $r=r_p+\delta r e^{\lambda t},\ n=n_p+\delta n e^{\lambda t}$ (with $n_p=\ron/(\ron+r_p)$) yields an equation for the dimensionless growth rate $\bar\lambda=\lambda/\koffo$ (Supplementary  Section S3) :
 \be
\bar\alpha_1\frac{n_p}{r_p}+(\log{r_p}+1)\left(1+\frac{\bar k_{\sigma}}{\bar\lambda}\right)=\frac{n_p\bar\alpha_1\log{r_p}}{(\bar\lambda+r_p+r_{\rm on})}.
\label{growthrate}
 \ee
 where the dimensionless  stiffness $\bar k_\sigma=k_\sigma/(\alpha_0\koffo)$ compares the dynamics of cell tension variations to the kinetics of linkers binding and unbinding.
 The system undergoes a supercritical Hopf bifurcation (${\rm Re}(\lambda)>0$), leading to a stable limit cycle \cite{guckenheimer:2002}, when
 \be
\frac{\bar\alpha_1}{\bar\alpha^*_{1,0}}-1> \frac{\bar k_\sigma}{\ron+r_p}\ \ ;\ \ \bar\alpha^*_{1,0}=\frac{r_p(r_p+\ron)^2(\log{r_p}+1)}{\ron(r_p\log{r_p}-r_p-\ron)}
 \label{stab_elast}
 \ee
If $\bar k_\sigma\rightarrow0$, this condition means that the fixed point lies in the unstable branch of the force/velocity curve ($v_{\rm slip}<v_p<v_{\rm stick}$ in \fig{fig2_protrusion}b). The stick-slip range is reduced for finite values of $\bar k_\sigma$ (\fig{fig2_protrusion}a). If \eq{stab_elast} is satisfied, the elastic protrusion undergoes permanent oscillations, alternating between  phases of growth (high friction and low retrograde velocity) and retraction (low friction and high retrograde velocity). 
The oscillation period depends on the difference between the slipping and the sticking forces and thus increases with the substrate adhesion strength $\bar\alpha_1$ (Supplementary Fig.S2).
Under physiological conditions, for which we estimate  $\ron\simeq10$, $v_\beta=1\unit{nm/s}$, $\alpha_0v_\beta=0.1\unit{\mu N/m}$,  $\bar\alpha_1\simeq 10^3$ and $\bar k_\sigma \simeq 1$  (Supplementary Table S.1), stick-slip is expected for a broad range of polymerisation velocity $v_{\rm slip}\simeq 15\unit{nm/s}<v_p<v_{\rm stick}\simeq 270\unit{nm/s}$. The typical slipping force $f_{\rm slip}\simeq 0.1\unit{mN/m}$ is consistent with traction force measurements \cite{gardel:2008} and with the tension of crawling cells.

\underline{\em Viscoelastic protrusion.} At long time ($t\gg\tau_\sigma$), the tension is that of a viscous fluid: $\sigma=\sigma_0+\alpha_\sigma \partial_t L$ with $\alpha_\sigma=k_\sigma\tau_\sigma$, and the protrusion's average length varies linearly with time. The short-time elastic response, including the stick-slip instability described above, nevertheless persists if tension relaxation is  slower than the linkers kinetics ($\koffo\tau_\sigma>1$ - Supplementary Section S3). The protrusion then alternates phases of growth and retraction with an average net growth (\fig{fig2_protrusion}c). 
Oscillations of the cell edge are observed in a limited range of values for the homeostatic cell tension $\sigma_0$. Indeed, as for elastic protrusions, stick-slip of visco-elastic protrusions requires that the steady-state retrograde velocity lies within the decreasing branch of the force-velocity curve (\fig{fig2_protrusion}b). This velocity can be modulated by factors affecting $\sigma_0$, in particular as acto-myosin contraction (\fig{fig2_protrusion}d). 

\underline{\em Viscoelastic protrusion; Kelvin-Voigt model.} 
For very large strain, membrane reservoir should eventually become unable to provide the membrane area needed to regulate membrane tension leading to an elastic behaviour with a different stiffness $k'_\sigma$ and to a saturation of the protrusion length. This can be accounted for by the so-called Kelvin-Voigt viscoelastic model: $\sigma=\sigma_0+k'_\sigma L+\alpha_\sigma \partial_t L$. This is dynamically equivalent to a purely elastic protrusion with a larger bare friction parameter $\alpha'_0=\alpha_0+\alpha_\sigma$. For single protrusions, the bare friction coefficient may thus be thought of as including viscous dissipation within the cell itself. 
The stability threshold \eq{stab_elast} is still valid, but with $\bar\alpha_1=\alpha_1/\alpha_0'$ and $\bar k_\sigma=k'_\sigma/(\alpha_0'\koff0)$.

\begin{figure*}[t]
\centerline{\includegraphics[width=17.8cm]{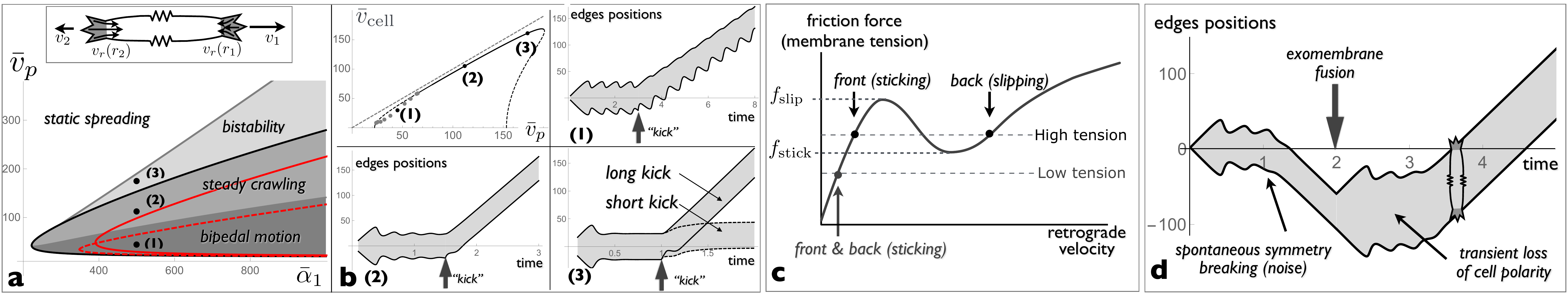}}
\caption{{\em Polarisation and crawling of a 1D elastic cell} {\bf (a)} Top: sketch of a 1D cell with protrusions are both ends coupled through the elastic cell tension (springs with a dimensionless stiffness $\bar k_\sigma$). Bottom:  crawling phase diagram with the adhesion parameter $\bar\alpha_1$ and the polymerisation velocity $\bar v_p$.
The symmetric spreading state is oscillatory within the thick solid lines (\eq{stab_elast}, black $\bar k_\sigma\rightarrow0$ and red for $\bar k_\sigma=10$). 
A crawling state exists within the regions shaded gray. Three regions can be distinguished: a  steady crawling state, where the cell moves without changing its shape, coexists with the static spreading state in the {\em bistability} region, and with the oscillatory spreading state in the {\em steady crawling} region. In the  {\em bipedal motion} region, no true steady-state exist and a bipedal crawling state - where the cell leading and trailing edges follow different limit cycles - coexist with the oscillatory spreading state. The dashed red lines is the boundary between steady crawling and bipedal motion for $\bar k_\sigma=10$.
{\bf(b)} Top: variation of the cell velocity with the polymerisation velocity. The solid  part of the curve correspond to steady crawling and the dashed parts are unstable steady-states. Gray dots correspond to bipedal motion. The other panels are examples of ``kymographs'' showing  the position of the two cells ends as a function of time in the three different regimes. The cell is first allowed to spread isotropically and is given an asymmetric ``kick'' which transiently removes  all bound linkers on one side of the cell at a prescribe time (arrows). In the bistable region, binder may rebind immediately after the kick (short kick) or after a short delay $\Delta t=0.05/\koffo$ (long kick). {\bf (c)} Role of membrane tension on cell polarity. The coexistence of a sticking and a slipping regime, necessary for cell polarisation,  requires high enough tension. The slipping state is inaccessible if membrane tension is decreased. {\bf (d)} Cell edge position as a function of time. Isotropic spreading starts at $t=0$, and is followed by a spontaneous breaking of symmetry due to intrinsic noise on $\ron$. Membrane tension is abruptly halved at $ t=2/\koffo$,  {\em e.g.} following the fusion of exocellular vesicles, resulting in a transient loss of cell polarity. High tension is restored after further spreading, leading to a new symmetry breaking event.
Parameters: $\ron=10$, {\bf (b)}: $\bar\alpha_1=500$ and $\bar k_\sigma=10$, {\bf (d)}:, $\bar v_p=100$ and $\ron$ is a gaussian random variable varying within $1\%$ around $\ron=10$.
\label{crawling}}
\end{figure*}

\subsection{Symmetry breaking and crawling of an elastic cell}
We  consider the dynamics of a simple 1D cell with a polymerisation-driven protrusion at the two cell ends (\fig{crawling}).  
In a first instance, we assume a uniform elastic cell tension  so that the two cells ends experience an instantaneous mechanical coupling and feel the same tension. The dimensionless off-rates $r_i$ and the fractions of bound linkers $n_i$ at the two cell ends ($i=\{1,2\}$) satisfy the equations:
  \begin{eqnarray}
\frac{\partial  n_i}{\partial \bar t}=\ron-(\ron+r_i)n_i \ ;\  \left(1+\bar\alpha_1\frac{n_i}{r_i}\right)r_i\log r_i=\bar\sigma\cr
\frac{\partial \bar\sigma}{\partial \bar t}=\bar k_\sigma (2\bar v_p-r_1\log r_1-r_2\log r_2)\hspace{1.5cm}
\label{elastic_crawling}
\end{eqnarray}
with $\bar\sigma=\sigma/(\alpha_0 v_\beta)$ and $\bar v=v/v_\beta$. The cell size satisfies $\partial_t L=v_1+v_2$, and the cell velocity is $v_{\rm cell}=(v_1-v_2)/2$. Static spreading corresponds to the retrograde velocity matching the polymerisation velocity at both ends ($v_1=v_2=0$, or $r_1=r_2=r_p$). Two independent modes of fluctuation exist around this state. The symmetric mode, where both ends  protrude and retract in synchrony ($v_1=v_2$), and the anti-symmetric mode, where one end retracts while the other protrudes ($v_1=-v_2$), leading to net cell translocation. The former is akin to two symmetric elastic protrusions and is unstable under the condition given by \eq{stab_elast} with $k_\sigma\rightarrow 2k_\sigma$. 
The latter occurs without membrane stretching and corresponds to $k_\sigma=0$.

In the stick-slip regime, the local force balance $\ff=\sigma$ admits three solutions, corresponding to the three branches of the force-velocity curve (labelled sticking branch, unstable branch and slipping branch in \fig{fig2_protrusion}b). For an elastic cell with a uniform tension, a steady crawling state exists in a limited range of polymerisation velocity where the retrograde velocities $v_{r,1}$ and $v_{r,2}$ at the two cell ends are on different branches and satisfy $\ff(v_{r,1})=\ff(v_{r,2})$ and $v_{r,1}+v_{r,2}=2v_p$. The conditions for the existence of such solution, together with  the stability boundary for static spreading (\eq{stab_elast}), lead to the crawling phase diagram shown in \fig{crawling}a. 
Linear stability analysis (Supplementary Section S4 and Fig.S5) show that steady crawling with the cell front  on the sticking branch and the rear on the slipping branch of the $\ff(v_r)$ curve is always stable. It coexists with a stable static  state in the ``bistability'' region and with an oscillatory  spreading state in the ``steady crawling'' region of the phase diagram.
For smaller values of the polymerisation velocity, steady crawling corresponds to the cell front on the sticking branch and the rear on the unstable branch of the $\ff(v_r)$ curve. This state is unstable if $\bar k_\sigma\rightarrow 0$, and one or both cell ends follow a limit cycle, resulting in an unsteady motion with a delay between protrusion of the front and retraction of the rear. This is the ``bipedal'' region  of the phase diagram, in reference to the bipedal motion of keratocytes \cite{barnhart:2010} which has been discussed in earlier theoretical works \cite{barnhart:2010,loosley:2012,ziebert:2013,camley:2013}.  Bipedal motion always coexists with the oscillatory spreading state. 
The boundary between steady crawling and bipedal motion depends on the value of the dimensionless cell stiffness $\bar k_\sigma$ as shown in \fig{crawling}a). 

Spontaneous cell polarisation and crawling can result from intracellular noise (for instance on the binding rate $\ron$, \fig{crawling}d), or can be triggered externally.
\fig{crawling}b shows the result of a ``kick'', a transient increase of the retrograde force leading to complete linkers detachment at one end of the cell.  In the ``steady crawling'' and ``bidepal'' phases, a  short kick applied to the oscillatory symmetric state is often sufficient to elicit cell polarisation and motion. The timing of the kick has an impact and symmetry breaking is triggered more efficiently if the kick  is applied during the spreading rather than the retracting phase of the symmetric oscillatory cycle.
Close to the bistability boundary, a short kick leads to small cell translocation but permanent polarisation requires a longer kick. Examples of these behaviours, together with the relationship between cell velocity and the polymerisation velocity in the different crawling regimes, are shown in \fig{crawling}b.


 The cell tension $\sigma$ plays an important role in cell polarisation. The existence of a motile state requires that the tension is sufficiently high to access the unstable branch of the force-velocity curve: $f_{\rm slip}<\sigma<f_{\rm stick}$ (see \fig{crawling}c). Such level of tension is naturally reached during  spreading if $v_p>v_{\rm slip}$, allowing the cell to spontaneously polarise and crawl (see \fig{crawling}d in the presence of intracellular noise). If the membrane tension of the crawling cell is abruptly decreased below $f_{\rm slip}$, {\em e.g} after the fusion of extracellular vesicles with the cell membrane as in  \cite{lieber:2013}, the slipping state is temporarily unaccessible and the cell becomes unpolarised, with both ends in the sticking regime. The cell then spreads further and membrane tension increases, eventually triggering a new event of spontaneous cell polarisation \fig{crawling}d. The steady-state  tension is thus entirely determined by the polymerisation velocity and the adhesion strength, and not by the amount of available membrane area, in agreement with experimental observations \cite{lieber:2013}.
 
 An alternative model in which the linear viscous force at high retrograde flow speed (parameter $\alpha_0$) is due to the viscoelasticity of the cytoskeleton rather than to substrate friction is studied in the Supplementary Section S5 (Fig.S6). In this case, the unstable branch persist up to $v_r\rightarrow\infty$ and the bistability region disappears.

\subsection{Persistent tension gradients}
In the elastic model  above,  propagation of mechanical stress across the cell is instantaneous, and  the cell tension is uniform.
In moving keratocytes, membrane tension is typically higher at the front than at the back of the cell \cite{lieber:2015}. The cell membrane being fluid, persistent tension gradients are necessarily generated by viscous dissipation - {\em e.g.} through friction between the cell membrane and the substrate \cite{schweitzer:2014,fogelson:2014}.  Membrane tension is thus a local quantity, characterised by  two parameters: an elastic stiffness $k$ - the local equivalent of the global stiffness $k_\sigma$ with the expected scaling $k_\sigma=k/L$ where $L$ is the cell length - and a local friction coefficient $\zeta$.  
Calling $u(z)$ is the local membrane displacement and combining the constitutive relationship $\sigma_m=k\partial_z u$ with the local membrane force balance: $\zeta \dot u=\partial_z \sigma_m$, yields a diffusion equation for the membrane tension:
\be
\partial_t\sigma_m(z)=\frac{k}{\zeta}\partial_z^2\sigma_m(z),
\label{difftension}
\ee 
with a diffusion coefficient $D=k/\zeta$. A $20\%$ difference between the front and rear membrane tensions of keratocytes crawling at a velocity $v\simeq0.1\unit{\mu m/s}$ \cite{lieber:2015} yields an estimate of  $\zeta\simeq10\unit{Pa.s/\mu m}$ and $D\simeq1\unit{\mu m^2/s}$ (Supplementary Table S1). Diffusive propagation of mechanical tension is also expected within the cytoskeleton due to its poro-elastic nature,  with similar values for the diffusion coefficient \cite{charras:2005}.  \eq{difftension} may thus be assumed to describe the spatio-temporal variation of the total tension $\sigma=2\sigma_m+\sigma_c$.

\begin{figure}[t]
\centerline{\includegraphics[width=8.7cm]{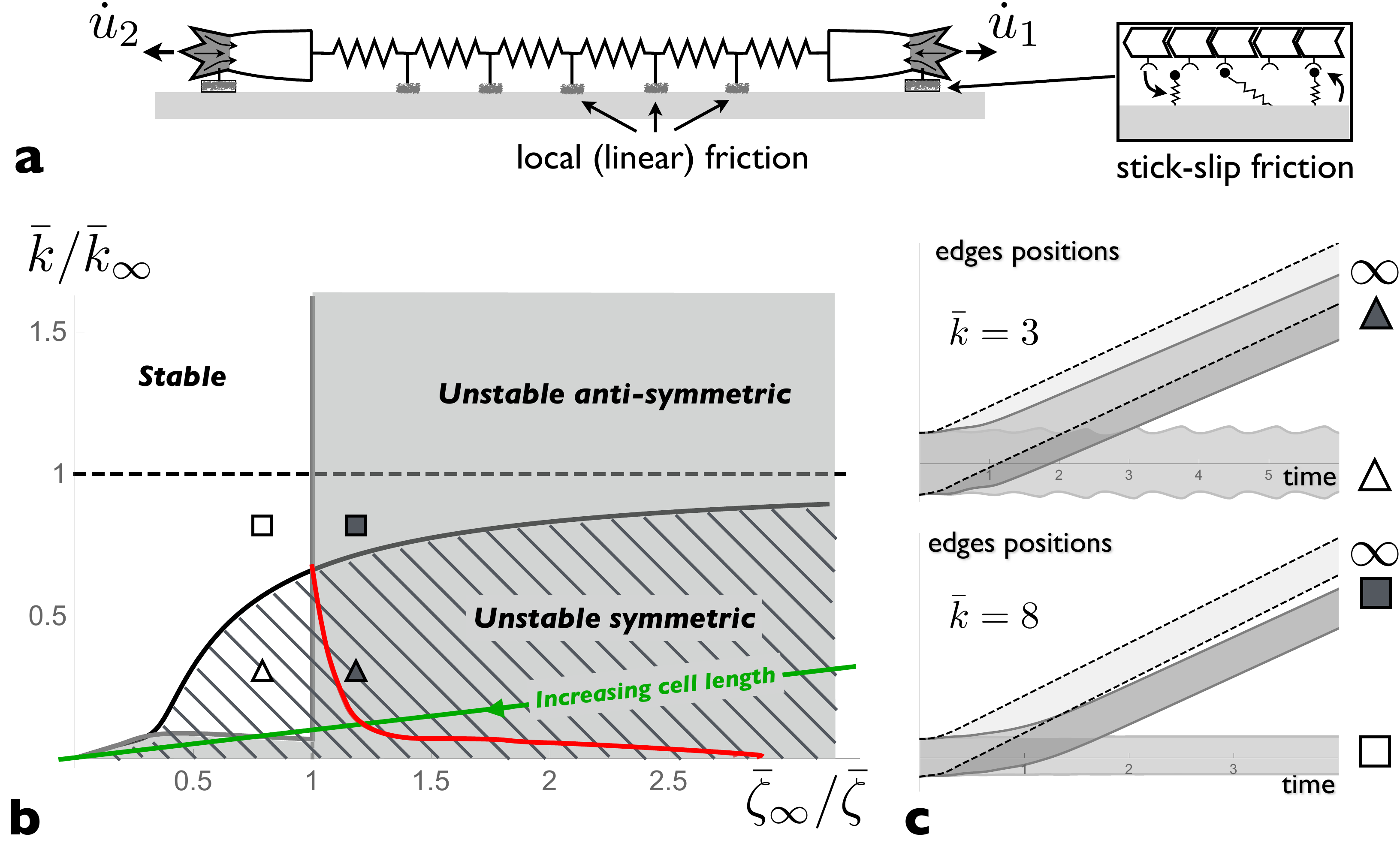}}
\vspace{2mm}
\caption{{\em Symmetry breaking with tension gradients.} {\bf (a)} Sketch of a 1D cell with protrusions at both ends, coupled by elastic elements experiencing friction from the substrate. {\bf (b)} Stability phase diagram of static spreading with the inverse substrate friction $1/\bar\zeta=\alpha_0/(\zeta L)$ and stiffness $\bar k=k/(L\alpha_0\koffo)$, normalised by their asymptotic values for $\bar D\rightarrow\infty$ (\eq{asymptotics}). Parameters  are such that the $\bar\zeta\rightarrow0$ limit is in the steady-crawling region of \fig{crawling}a. 
A small perturbation (the removal of  $10\%$ of the bound linkers at one end of a cell initially at the symmetric stationary state) leads to permanent cell polarisation and crawling above the red line. Below this line, the end state is symmetric. The green line shows the effect of increasing the cell length, all other dimensional parameters being unchanged.
{\bf (c)} Example of cell trajectories for the parameters given by the different symbols in {\bf (b)}. Dashed lines are the trajectories for $\bar D\rightarrow\infty$.  For $\ron=10,\ \bar\alpha_1=500,\ \bar v_p=100$ (corresponding to $\bar k_\infty\simeq 10$ and $\bar\zeta_\infty\simeq 1$).
 \label{fig_diffusion}}
\end{figure}

Tension gradients generated by periodic protrusion/retraction cycles of a cell edge with a period $\tau$ (of order $20\unit{s}$ in \cite{giannone:2007}) decay over a length scale $\sqrt{D\tau}\simeq   5\unit{\mu m}$. Cell edges much further apart are essentially mechanically independent. 
To account for this,  the tension appearing in the force balance equation in \eq{elastic_crawling} must be replaced by the tension at either end of the cell $\sigma(\pm L/2)$ calculated with \eq{difftension}, supplemented with the boundary conditions $\dot u_{1,2}=\dot u(\pm L/2)=\pm(v_p-v_{r(1,2)})$ (\fig{fig_diffusion}a). 
Linear expansion of the dynamical equations (Supplementary Section S6) show that the growth rate equation \eq{growthrate} still holds, albeit with  rate-dependent effective stiffnesses $\bar k^\pm_{\sigma,\lambda}$ for symmetric ($+$)  and anti-symmetric ($-$)  perturbations  given by:
\begin{eqnarray}
\bar k_{\sigma,\lambda}^\pm=\bar k\sqrt{\bar\lambda/\bar D}\left(\coth{\sqrt{\bar\lambda/\bar D}}\pm \frac{1}{\sinh{\sqrt{\bar\lambda/\bar D}}}\right),
\label{kdiffspreading}
 \end{eqnarray}
with $\bar k=k/(L\alpha_0\koffo )$, $\bar \zeta=\zeta L/\alpha_0$, $\bar\lambda=\lambda/\koffo$, and $ \bar D=\bar k/\bar\zeta$. 

Linear stability analysis of static spreading must be performed  numerically and leads to the phase diagram shown in \fig{fig_diffusion}b. 
Asymptotic behaviours are derived in Supplementary Section S6. In the limit of high friction: $\bar D\ll1$, tension variation near the tip of a moving protrusion relax over a very short length scale $\sqrt{D|\lambda|}\ll L$, leading to a large effective stiffness  $\bar k_{\sigma,\lambda}\sim\sqrt{\bar k\bar\zeta/|\bar\lambda|}$. Protrusions are unstable below a threshold effective stiffness, leading to the scaling $\bar k\sim 1/\bar\zeta$ for the stability boundary of both types of perturbations. 
In the limit of low friction: $\bar D\gg1$, symmetric perturbations are unstable below a threshold stiffness $\bar k_\infty$
and anti-symmetric perturbations are unstable below a threshold friction $\bar \zeta_\infty$, given by:
\be
\bar k_\infty=\frac{\ron+r_p}{2}\left(\frac{\bar\alpha_1}{\bar\alpha_1^*(0)}-1\right)\ ,\  \bar \zeta_\infty=2\left(\frac{\bar\alpha_1}{\bar\alpha_1^*(0)}-1\right)
\label{asymptotics}
\ee
with $\bar\alpha^*_{1,0}$ given by \eq{stab_elast}. 

In regions of the phase diagram where both the symmetric and anti-symmetric perturbations are unstable, the end state may be the polarised crawling state or  the oscillatory spreading state, depending on initial conditions. 
Under high friction the stability criterion is the same for symmetric and anti-symmetric fluctuations because the two protrusions are independent. This means that fluctuations are unlikely to lead to persistent cell polarisation. 
Numerical solutions of the coupled equations for linkers kinetics and diffusion of the cell tension are shown in \fig{fig_diffusion} for cells initially in the static spreading state with $10\%$ of bound linkers removed at one end of the cell - a moderate ``kick'' compared to \fig{crawling}b. This asymmetric kick leads to persistent polarisation and crawling if the dimensionless friction coefficient $\bar\zeta=\zeta L/\alpha_0$ is below a threshold value that depends on the dimensionless stiffness. If the lamellipodium size is uncorrelated with the cell size, so that both $\alpha_1$ and $\alpha_0$ are independent of $L$, polarisation is predicted to occur below a threshold cell size (\fig{fig_diffusion}b). 

 \subsection{Travelling lateral waves} 
\begin{figure}[t]
\centerline{\includegraphics[width=8.7cm]{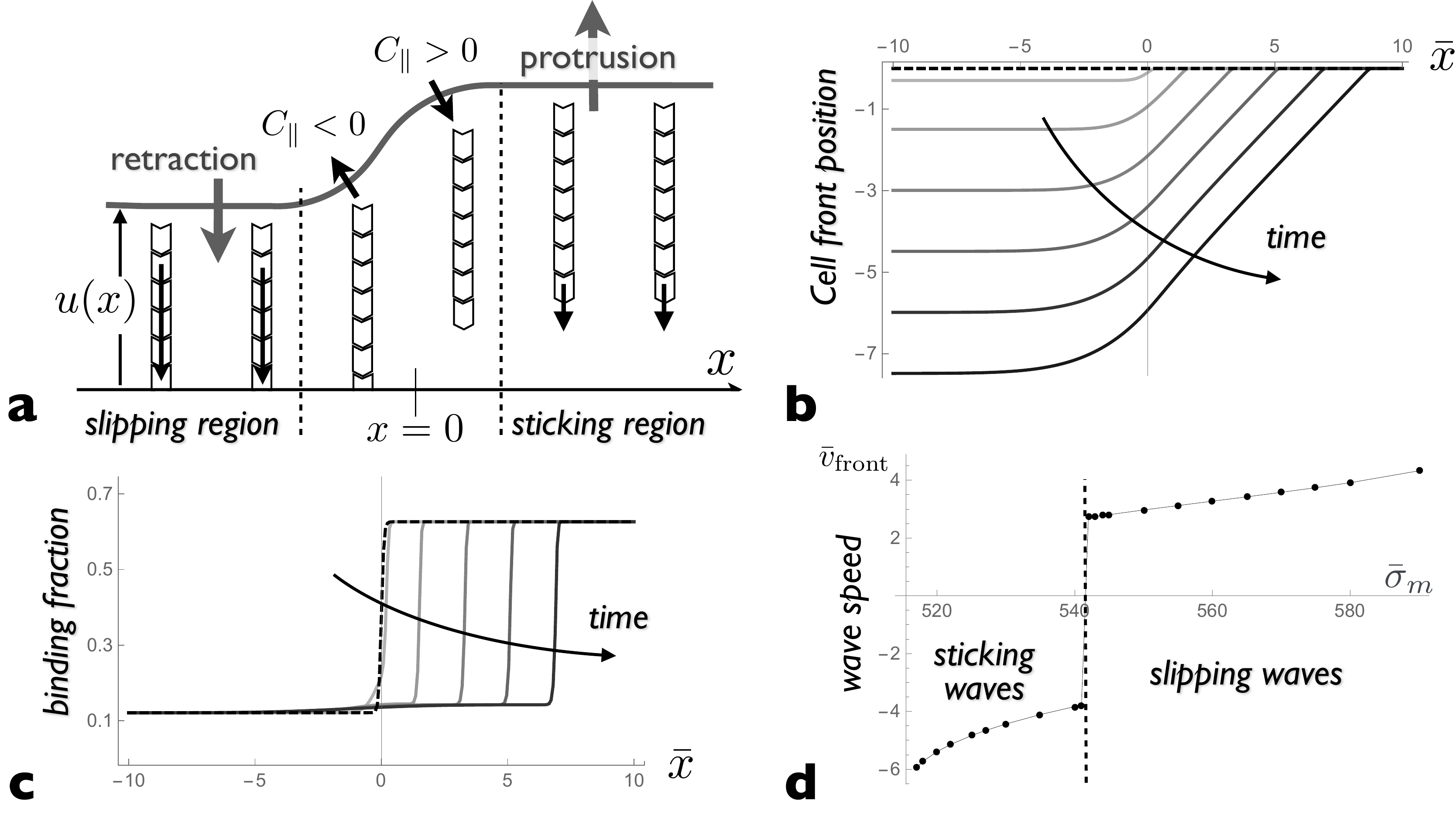}}
\caption{{\em Travelling wave} {\bf (a)} Sketch of the cell edge at an  interface between a region of fast retrograde flow (slipping) and a region of slow retrograde flow (sticking). The edge curvature modify the force driving actin retrograde flow. Positive (negative) curvature reduces (increases) the driving force, triggering a transition between the sticking and slipping states, and the lateral motion of the interface. {\bf (b)} Evolution of the cell edge profile relative to the protrusive side  with time, and {\bf (c)} evolution of the fraction of bound linkers with time (for equal time steps up to $t=2.8/\koffo$). A travelling wave moves at a constant velocity, toward the sticking region (slipping wave) in this particular example (examples of sticking waves are shown in Supplementary Fig.S7). 
{\bf (d)} The front velocity depends on membrane tension, with an abrupte transition between a sticking wave ($v_{\rm front}<0$) and a slipping wave ($v_{\rm front}>0$) observed at a particular value of the membrane tension. Parameters: $\ron = 10, \bar\alpha_1 = 500,\ \bar v_p = 100$ and $\bar\sigma_m = 285$.
\label{travelling}}
\end{figure}

Travelling waves are ubiquitous in motile cells \cite{giannone:2004,Dobereiner:2006,dubinthaler:2008}, and
have been discussed in the context of the active mechanics of the cytoskeleton or the reaction-diffusion dynamics of  biochemical cytoskeleton regulators
 \cite{allard:2012}. The abrupt nature of the switch between high and low friction states suggests the possibility for travelling waves of purely mechanical origin.
To see this, we extend the model to a 2D cell edge under small deformation. Let's  consider an initially flat cell edge where regions in the sticking and slipping state coexist. In \fig{travelling}a, half the edge ($x>0$) is in the sticking state and the other half ($x<0$) in the slipping state at $t=0$, where $x$ is the coordinate along the cell edge. As the spreading velocity is different in these two regions, a kink forms and grows at the boundary between them, with  a positive curvature in the sticking region and a negative curvature in the slipping region. Calling $u(x)$ the position of the cell front, and assuming small deformation ($\partial_x u\ll1$), the membrane force in \eq{forcebal} reads $2\sigma_m (1+ h C_\|)\simeq 2\sigma_m (1- h \partial^2_x u(t))$. The positive (negative) curvature increases (decreases) the retrograde force. Since the curvature increases with time, one side of the boundary  between the two states (or sometimes both, Supplementary Section S7 and Fig.S7) eventually undergoes the stick-slip transition leading to a lateral movement of the boundary.

At lowest order in edge deformation, the evolution of the edge profile is given by $\partial_t u(x)=v_p-v_\beta r(x)\log{r(x)}$, with a position-dependent retrograde velocity obtained from Eqs.(\ref{forcebal},\ref{ff1}):
\be
\frac{\partial r(\bar x)}{\partial\bar t}=\frac{\partial_{\bar x}^2(r(\bar x)\log{r(\bar x)})-\bar\alpha_1\log{r(\bar x)}\dot n(\bar x)}{\log{r(\bar x)}+1+\bar\alpha_1n(\bar x)/r(\bar x)}
\label{travellingeq}
\ee
with the local linker's kinetics still given by \eq{ff1}. Here, the dimensionless spatial coordinate is $\bar x=x\sqrt{\alpha_0\koffo/(2\sigma_m h)}$.
These equations supports travelling waves in the stick-slip regime, as can be seen from the evolution of the edge profile and the density of bound linker (\fig{travelling}b.c). The front velocity scales as $v_{\rm front}=\sqrt{2\sigma_m h\koffo/\alpha_0}\bar v_{\rm front}(\bar\sigma_m,\bar\alpha_1,\ron)$, where $\bar v_{\rm front}$ is a dimensionless function of dimensionless parameters. \fig{travelling}d shows how $\bar v_{\rm front}$ depends on the membrane tension $\bar\sigma_m$. The most remarkable feature is a transition from a sticking wave to a slipping wave (a change of sign of $v_{\rm front}$) above a threshold tension. This is because the slipping state is more stable at high tension. The dependence with the two other parameters is less interesting (Supplementary Section S7). Using physiological parameters (Supplementary Table S1), $v_{\rm front}\sim \unit{0.1\mu m/s}$, which agrees with the order of magnitude observed in mouse embryonic fibroblasts and T cells \cite{Dobereiner:2006}. 

 \section{Discussion}
 
Many of the molecular players involved in cell motility have been identified, but we are still searching for the basic principles underlying their organisation in space and time. 
A growing body of evidence shows that mechanics plays a key role in organising cell motility, be it the active mechanics of the cytoskeleton, the tension of the cell membrane, or  the stiffness of the substrate. We propose a simple model of cell spreading and crawling based on the interplay between actin polymerisation and  cell-substrate adhesion mediated by mechano-sensitive stochastic linkers. Despite its simplicity, the model reproduces a number of cellular behaviours reported in the literature: (i) the stick-slip behaviour of the cell front \cite{giannone:2004,giannone:2007}, (ii) spontaneous symmetry breaking or bistability of cells and cell fragments \cite{Verkhovsky:1999,yam:2007}, 
(iii) the crucial role of membrane tension in regulating this process \cite{lieber:2013}, and (iv) the existence of slipping or sticking lateral waves propagating along the cell edge \cite{giannone:2004,Dobereiner:2006,dubinthaler:2008}. 
The model also uncovers the few dimensionless parameters controlling the transition between different  dynamical behaviours.

A number of simplifying assumptions are made regarding the mechanics and dynamics of the cytoskeleton. 
The rate of actin polymerisation and the distribution of acto-myosin contractile stress, assumed constant and uniform here, do vary across the cell.
Transient loss of adhesion can locally increase the actin density and lead to the formation of ruffles or actin arcs stabilised by acto-myosin contraction \cite{burnette:2011}. 
As discussed in the introduction, modulation of these parameters through feedback loops can also lead to some of the dynamical features explained here.
It is nevertheless important to study individually the different modules that can be combined to regulate the behaviour of crawling cells. The present simplifications allows to precisely focus on the role of mechano-sensitive adhesion.

The dynamics of cellular protrusions driven by actin polymerisation result from a balance between  acto-myosin contraction and  membrane tension, which generate an actin retrograde flow, and the traction force generated by stochastic adhesion bonds transiently linking actin filaments with the substrate. 
Collective effects within the population of mechano-sensitive bonds lead to a non-monotonic relationship between the steady-state traction force and the retrograde flow velocity (\fig{fig2_protrusion}b). This defines a range of polymerisation velocity within  which protrusions may exhibit a stick-slip dynamics, alternating phases of growth with a weak retrograde flow and retraction with a strong retrograde flow. Whether such instability develops depends on the way the tension at the tip of a protrusion varies with the protrusion length and growth rate. For simple elastic cells with a uniform tension, 
this is characterised by a dimensionless effective stiffness $\bar k_\sigma=k_\sigma/(\alpha_0\koffo)$, which compares the rates of variation of the cell tension and of the density of attached bonds. 
Tension variations during growth and retraction attenuate the collective effects among the adhesion bonds, so that single protrusions display a stick-slip dynamics if  $\bar k_\sigma$ is below a threshold value, that is if the cell (and in particular the cell membrane) is sufficiently soft.

For purely elastic protrusions, the protrusion length averaged over the periods of growth and retraction remains constant.
If the cell tension is viscoelastic with a long-time viscous behaviour, the short-time periodic stick-slip behaviour superimposes on a slower long-time growth (\fig{fig2_protrusion}c,d), leading to a dynamics that strongly ressembles, both in shape and time scale (period $\sim 20\unit{s}$), the periodic oscillations of the edge of spreading fibroblasts \cite{giannone:2004,giannone:2007}. 
The periodic buckling of the lamellipodium proposed in \cite{giannone:2007} to explain these observations is not incompatible with the present explanation. Indeed, buckling, along with actin arcs and ruffles, can constitute a non-linear response to the abrupt actin unbinding from the substrate, which is at the origin of the oscillations proposed here. 
Importantly, oscillations of the leading edge are observed within a finite range of cell homeostatic tension (\fig{fig2_protrusion}d). They can be suppressed by Myosin inhibition, as observed in \cite{giannone:2004,giannone:2007}, but also by myosin over-expression, which is a falsifiable prediction of the present model.
The oscillatory behaviour of cellular protrusions can also be modulated by factors regulating membrane tension, such as microtubules, SNARES, or dynamin. 

Spontaneous cell polarisation and crawling is studied using a simple 1D model where two polymerisation-driven protrusions form at both ends of the cell and are mechanically coupled by the cell tension. This simplified geometry allows to disregard the complex interplay between the cell shape and the local force balance in 2D motility \cite{keren:2008}. It  is  appropriate for cells confined on adhesive tracks \cite{maiuri:2012}, and is also relevant for  motility in physiological 3D matrices, which appears closer in many way to 1D than to  2D motility \cite{Doyle:2013}.

In addition to the local balance between the traction force and the cell tension at each cell end, 
the net force exerted by the cell on the substrate must vanish by virtue of Newton's third law. This is naturally satisfied for a symmetric cell, which can either settle into a static spreading state where the retrograde flow compensates the polymerisation velocity at both end, or into an oscillatory  state alternating phases of symmetric spreading and retraction around a fixed length, without net cell translocation. Crucially, the non-monotonic nature of the force-velocity relationship allows for a state of broken symmetry, in which the cell is polarised and motile with the leading edge in the sticking state and the trailing edge in a slipping state.
Depending on the parameters, the crawling state can be a true steady-state where the cell moves without changing its shape, or an unsteady moving state where  the leading and trailing edges display asynchronous oscillation reminiscent of the bipedal motion of keratocytes \cite{barnhart:2010}.  

If the cell tension equilibrates fast and is uniform, the motility behaviour is mostly determined by the polymerisation velocity and the strength of adhesion (\fig{crawling}). 
Under low adhesion, the cell is in the static spreading state and cannot crawl. Upon increasing the adhesion strength, a region of bistability emerges, where the transition from a (meta)stable static state and a steady crawling state needs to be triggered by large  fluctuations or external ({\em i.e.} mechanical) perturbations. For larger adhesion strength, the static state is unstable and the polarised state competes with an oscillatory spreading state.
Finally, for even larger adhesion, crawling is unsteady and the leading and trailing edges of the cell display asynchronous oscillations. 
The value of the effective cell stiffness $\bar k_\sigma$ modifies the boundaries between static and oscillatory spreading and between steady and bipedal crawling (\fig{crawling}). A large effective stiffness, which corresponds to a small cell length, increases the bistability region. 
This could explain why small cell fragments are prone to bistability  \cite{Verkhovsky:1999}. Symmetry breaking is often explained by the redistribution of actomyosin contraction to the back of the cell \cite{Verkhovsky:1999,yam:2007,hawkins:2011,ziebert:2012,tjhung:2012,ruprecht:2015,Maiuri:2015,kozlov:2007}, and myosin activity does increase the probability of symmetry breaking and the velocity of cells and cell fragments in 2D\cite{Verkhovsky:1999,yam:2007}. It is noteworthy however that a majority of keratocyte fragments in \cite{Verkhovsky:1999} are polarised and motile even in the presence of drugs inhibiting myosin activity. This suggests that the mechano-sensitive adhesion switch described here can be sufficient to elicit the excitable behaviour of crawling cells, which is enhanced through the feedback between actin flow and myosin distribution. 

The model identifies the cell tension, and in particular the tension of the cell membrane,  as key to the coordination between the two cell edges. Remarkably, a sudden decrease of membrane tension provoked by 
the fusion of extracellular vesicles leads to the formation of multiple lamellipodia that significantly hampers the cell's ability to polarise and crawl \cite{lieber:2013}. 
The present model recapitulates this behaviour, including the resumption of persistent polarisation and directed motion after further cell spreading, with tension values similar to those before vesicle fusion. Multiple lamellipodia form under low tension because the slipping state only exist under high enough tension, so that both ends of the cell are in a protrusive state under low tension. The homeostatic tension of crawling cells is determined by a balance between cytoskeletal forces and adhesion rather than by the available membrane area, in agreement with the conclusion of \cite{lieber:2013}. An increase of membrane tension upon cell spreading may thus  be the force driving the cell into bi-stability, allowing for the existence of a polarised, motile state.


While spontaneous symmetry breaking and motion upon spreading is common for cell fragment \cite{Verkhovsky:1999} and some cell types such as  keratocytes \cite{yam:2007}, other cells types  form and retract uncoordinated protrusions without global symmetry breaking. 
This is not easily explained if the cell tension is uniform
since cells with dynamic protrusions also possess a metastable crawling state (\fig{crawling}). The absence of symmetry breaking could be a matter of time scale, or related to the difference between 1D and 2D geometries. We propose instead that it is due to finite relaxation time of tension heterogeneities across the cell.  Membrane tension is larger at the front than at the rear of fast moving cells such as keratocytes  \cite{lieber:2015}, likely due to the existence of friction between the membrane and either the substrate, and element of the cytoskeleton \cite{schweitzer:2014,fogelson:2014}. As a consequence, tension relaxes in a diffusive manner, and dynamic protrusions further than a few $\mu m$ apart are mechanically independent. This fundamentally affects the crawling phase diagram (\fig{fig_diffusion}) and spontaneous symmetry breaking is limited to cells with low friction or equivalently to small cell size.

The extension of the present model to 2D geometry is not expected to alter our main conclusion. One interesting new feature that can emerge, however, is the propagation of lateral waves of protrusion/retraction along the cell edge  \cite{giannone:2004,Dobereiner:2006,dubinthaler:2008}. 
The lateral curvature of the cell edge affects the force on the actin filaments and can induce a stick-slip transition. We show that this leads to propagating waves at the interface between sticking and slipping regions at the edge of the cell, 
with a velocity ($\sim 0.1\unit{\mu m/s}$) comparable to that of lateral waves observed in MEF and  T cells \cite{Dobereiner:2006}. 
The velocity of these ``stick-slip waves'' is not controlled by the polymerisation velocity, but rather by a balance between membrane tension and substrate friction. Such waves constitute an alternative to  reaction-diffusion  waves of cytoskeleton regulators \cite{allard:2012} for rapid transmission of mechanical signals across the cell during migration.

To conclude, the stick-slip mechanism described here, which bestow motile cells with dynamical behaviours typical of excitable systems and can reproduce a diversity of experimental behaviour, is intimately linked to the interplay between the time scale of formation and disruption of cell adhesion and the visco-elastic and diffusive time scale of cell tension variation. Its full understanding requires a proper treatment of both dynamical processes such as the one proposed here. This adds a knob to the cell toolkit, that also includes diffusion/reaction of signalling molecules and  active cytoskeleton mechanics, to confer robustness and sensitivity to crawling cells.


\begin{acknowledgments}
I would like to thank Jacques Prost for a critical reading of the manuscript.
This work was partially supported by  the Human Frontier Science Program under the grant RGP0058/2011. 
\end{acknowledgments}



\end{document}